\documentclass[11pt]{article}

\usepackage[final]{acl}

\usepackage{times}
\usepackage{latexsym}

\usepackage[T1]{fontenc}

\usepackage[utf8]{inputenc}

\usepackage{microtype}

\usepackage{inconsolata}

\usepackage{graphicx}

\title{AI as Teammate or Tool? A Review of Human–AI Interaction \\in Decision Support}

\author{Most. Sharmin Sultana Samu\thanks{Corresponding author},  {\bf Nafisa Khan}, {\bf Kazi Toufique Elahi,} \\ {\bf Tasnuva Binte Rahman}, {\bf Md. Rakibul Islam}, {\bf Farig Sadeque} \\
        sharmin.sultana.samu@g.bracu.ac.bd, nafisa.khan@g.bracu.ac.bd, \\ kazi.toufique.elahi@g.bracu.ac.bd, tasnuva.binte.rahman@g.bracu.ac.bd, \\ md.rakibul.islam11@g.bracu.ac.bd, farig.sadeque@bracu.ac.bd \\
       \\ Department of Computer Science and Engineering, BRAC University, Dhaka-1212}

\begin{document}
\maketitle
\vspace{5pt}
\begin{abstract}
The integration of Artificial Intelligence (AI) necessitates determining whether systems function as tools or collaborative teammates. In this study, by synthesizing Human-AI Interaction (HAI) literature, we analyze this distinction across four dimensions: interaction design, trust calibration, collaborative frameworks and healthcare applications. Our analysis reveals that static interfaces and miscalibrated trust limit AI efficacy. Performance hinges on aligning transparency with cognitive workflows, yet a fluency trap often inflates trust without improving decision-making. Consequently, an overemphasis on explainability leaves systems largely passive. Our findings show that current AI systems remain largely passive due to an overreliance on explainability-centric designs and that transitioning AI to an active teammate requires adaptive, context-aware interactions that support shared mental models and the dynamic negotiation of authority between humans and AI.

\end{abstract}

\section{Introduction}

Artificial Intelligence has rapidly progressed from static automation systems to interactive Decision Support Systems (DSS) intended to augment human judgment in complex environments. In fields such as medical diagnosis, financial forecasting and public safety, the objective has shifted from replacing human experts to establishing hybrid intelligence, where humans and AI systems complement each other's strengths. This evolution raises a central question: \textbf{`Does the current generation of AI function as a collaborative teammate that shares context and goals, or does it remain a sophisticated tool requiring explicit, manual operation?'} The answer is rooted not in the underlying model architectures, but in the interaction interface, where algorithmic outputs intersect with human cognition.
\par
Despite advances in AI decision-making, many recent studies show significant blind spots. A portion of existing work focuses heavily on algorithmic performance metrics such as accuracy, F1 Scores and latency, while neglecting the human factors that determine real-world efficacy \cite{guerdan2023ground}. And where interaction is studied, it is often limited to short-term, single-session experiments that fail to capture how trust evolves \cite{sun2025revisiting}. Furthermore, recent studies highlight a transparency paradox: while explainable AI (XAI) is touted as a solution to trust issues, empirical evidence suggests that more information does not always lead to better decisions and can, in fact, increase cognitive load or induce over-reliance \cite{rezaeian2025explainability}. There is a lack of comprehensive reviews that connect these isolated findings on interface design, trust calibration, and collaborative frameworks into a cohesive narrative regarding the Teammate vs. Tool dichotomy \cite{bach2024systematic}.
\par
In this paper, we conduct a comprehensive review of the state of the art in Human-AI Interaction, explicitly focusing on literature published from 2023 to 2025. We move beyond simple performance evaluation to dissect the interaction layer of decision support. We have grouped the reviewed literatures into following parts: (i) Human–AI Interaction Design \& Interface Patterns, examining how visual and textual presentation shapes user behavior, (ii) Trust, Explainability and Human Confidence Calibration, analyzing the psychological mechanisms of reliance and skepticism, (iii) Collaborative Decision Frameworks \& Optimization of Human–AI Teams, reviewing system-level architectures for optimizing joint human-AI teams, and (iv) Human–AI Interaction in High-Stakes Healthcare Decision Support Systems, providing a deep dive into clinical environments where the consequences of interaction failures are most critical.
\par
In short, with this study, we have made the following contributions:
\begin{enumerate}
    \item A comprehensive overview of 24 recent works on human-AI interaction in decision support is provided.
    \item The literature is categorized into four distinct areas, with shortcomings and areas for improvement identified.
    \item Key findings are summarized, and emerging trends in the field are highlighted.
\end{enumerate}


\section{Human–AI Interaction Design \& Interface Patterns}
The design of interaction interfaces plays a central role in shaping how humans understand, trust and collaborate with AI systems in high-stakes domains. Prior work shows that performance gains from AI do not depend only on model accuracy. They depend strongly on how AI outputs, explanations and uncertainties are presented to users \cite{chen2025engaging, haque2024we, he2025conversational, cho2025human, morrison2023evaluating}. This taxonomy focuses on interface patterns that influence trust calibration, cognitive load, ethical judgment and decision quality. The literature reveals recurring trade-offs between engagement, transparency and appropriate reliance. It also exposes gaps in adaptive, long-term and context-aware interaction design.
    \subsection{Explanation and Transparency Interfaces}
    Several studies examine how explanation interfaces affect user understanding and reliance. Text-based and conversational explanations often increase perceived understanding and trust, but they do not reliably support better judgment \cite{chen2025engaging, he2025conversational}. Textual explanations improved engagement and perceived reliability in decision-support tasks, but they increased cognitive load and did not prevent users from following incorrect AI advice \cite{chen2025engaging}. Conversational XAI interfaces further amplified this effect. Interactive and LLM-based explanations increased confidence and trust, yet they strengthened over-reliance and reduced objective feature understanding, especially when explanations appeared fluent and human-like \cite{he2025conversational}.

    Visual explanation patterns show mixed results. Simple visual explanations encouraged passive acceptance and offered limited support for reflective reasoning \cite{chen2025engaging}. In contrast, spatial explanations grounded in task-relevant signals, such as Grad-CAM overlays for medical image registration, improved precision and specificity without increasing workload \cite{cho2025human}. Similarly, explanation strategies derived from human reasoning, such as causal explanations, helped users reduce over-reliance when AI predictions were incorrect, but only when localization quality was high \cite{morrison2023evaluating}. Across studies, explanation effectiveness depended on alignment between explanation type, task demands and AI correctness, rather than explanation richness alone.
    
    \subsection{Trust Calibration and Cognitive Load Management}
    Improper trust calibration emerges as a persistent challenge across domains. Users frequently over-relied on AI recommendations, even when AI performance was poor, leading to worse outcomes than AI-only baselines \cite{chen2025engaging, he2025conversational}. Interface mechanisms that exposed AI confidence supported better trust calibration by allowing users to compare their judgment with model certainty \cite{chen2025engaging, cho2025human}. However, mechanisms that increased reflection through questions or feedback often reduced performance and trust due to higher cognitive effort \cite{chen2025engaging}.

    Stakeholder studies in policing further show that trust is shaped by lived experience and domain expertise rather than interface transparency alone \cite{haque2024we}. Law enforcement agents often relied on initial system outputs despite high mental workload, while community members expressed skepticism rooted in ethical concerns. These findings suggest that trust calibration requires interfaces that reduce cognitive burden while preserving space for human judgment, rather than relying solely on richer explanations or interaction.
    
    \subsection{Interactive and Collaborative Interface Patterns}
    Beyond explanations, interactive visual systems can support sensemaking and exploratory reasoning. ThemeViz demonstrates that combining LLM assistance with structured visual overviews reduces cognitive load and supports iterative theme development better than prompt-based chat interfaces \cite{kang2025themeviz}. Centralized data management and grounded AI suggestions lowered hallucinations and improved usability. However, users still perceived the AI as a tool rather than a collaborator 

\begin{figure*}[t] 
    \centering
    \includegraphics[width=0.95\textwidth]{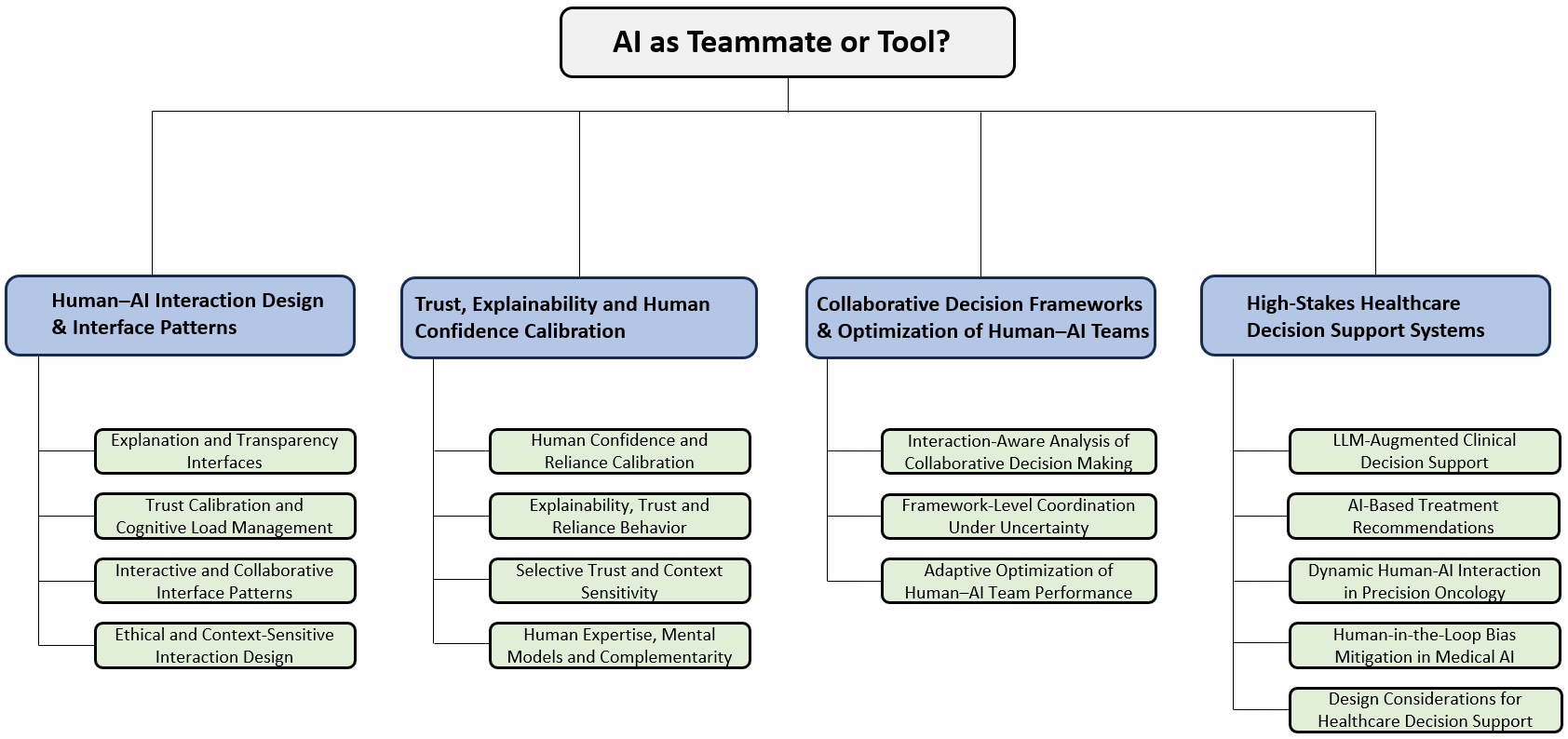} 
    \caption{Taxonomy of `AI as Teammate or Tool?'.}
    \label{fig:taxonomy}
\end{figure*}
    
    due to limited agency and one-directional interaction \cite{kang2025themeviz}. Similar patterns appear in other domains, where AI assistance improved task efficiency but did not achieve balanced collaboration or shared control \cite{cho2025human}.

    In safety-critical tasks, tightly integrated Human+AI and Human+XAI interfaces improved sensitivity, precision and specificity compared to human-only decisions \cite{cho2025human}. These gains occurred even when trust levels remained stable, indicating that effective collaboration can emerge from interface alignment without increased subjective trust. This contrasts with conversational interfaces, where higher trust did not translate into better outcomes \cite{he2025conversational}.

    \subsection{Ethical and Context-Sensitive Interaction Design}
    Ethical concerns strongly influence how users interpret and engage with AI systems. In policing contexts, interface design shaped perceptions of bias, over-policing and harm to marginalized communities \cite{haque2024we}. A mismatch between algorithmic constructs and domain concepts increased cognitive demand and risked reinforcing harmful practices. Similar concerns emerged in qualitative analysis tools, where users raised issues of data privacy, intellectual property and bias in black-box models \cite{kang2025themeviz}. These studies highlight that human-centered interface design must incorporate stakeholder values, domain semantics and ethical safeguards, rather than focusing only on usability or performance.

    \subsection{Dataset Design for Human–AI Interaction Studies}
    Several studies construct carefully controlled datasets to study human–AI decision-making across diverse domains, but they differ in task realism, data curation and control over AI behavior. In a healthcare setting, \cite{chen2025engaging} build a structured dataset for diabetes meal planning using a staged human–AI interaction design, where ground truth decisions are defined through nutritional rules and historical participant choices are reused to model AI confidence, enabling fine-grained analysis of explanation effects. In contrast, \cite{haque2024we} rely on large-scale real-world crime records and extensive preprocessing to support algorithmic visualization, prioritizing ecological validity over experimental control. Financial decision-making is addressed by \cite{he2025conversational}, which curate a public loan dataset to precisely control prediction correctness and confidence, ensuring a fixed AI accuracy that exposes participants to both reliable and misleading advice. Beyond decision accuracy, \cite{kang2025themeviz} focus on thematic interpretation tasks and assemble heterogeneous textual datasets from online discourse and historical interviews, emphasizing diversity and prompt feasibility rather than labeled correctness. For technical and visual domains, \cite{cho2025human} and \cite{morrison2023evaluating} construct highly curated datasets with strong ground truth, using simulation, filtering and augmentation to manage class imbalance and task difficulty in medical image registration and disaster damage assessment. Taken together, these datasets reflect a shared emphasis on careful preprocessing and task design, while revealing a trade-off between realism and experimental control that shapes how human–AI interaction outcomes can be evaluated.

    \subsection{Evaluation Metrics for Human–AI Interaction}
    Across studies, evaluation metrics combine objective performance with subjective user perceptions, but they differ in how deeply they capture trust, reliance and cognitive effects. Several works integrate accuracy-based measures with perception surveys to assess decision quality and user experience. In high-stakes decision tasks, \cite{chen2025engaging} and \cite{he2025conversational} jointly measure decision accuracy alongside trust, perceived reliability, confidence and system understanding, but \cite{he2025conversational} extends this by introducing reliance and appropriate reliance metrics such as Agreement Fraction and RAIR to detect over- or under-reliance on AI advice. In contrast, \cite{haque2024we} emphasize cognitive workload and task experience using NASA-TLX and qualitative interviews, prioritizing mental demand and frustration over direct decision accuracy. Visual and technical domains adopt more granular performance metrics. \cite{cho2025human} use classification measures such as sensitivity, precision and specificity to capture acceptance and rejection behavior, while also pairing them with workload and perceived understanding to study the role of explainability. Similarly, \cite{morrison2023evaluating} evaluate both assessment and localization accuracy, and explicitly measure behavioral reliance when AI outputs are wrong, which exposes how explanations shape corrective behavior. Beyond correctness, \cite{kang2025themeviz} focus on process-oriented metrics such as iteration counts, technology acceptance and perceived collaboration, supported by qualitative analysis of limitations. Together, these evaluation matrices reveal a common trend toward mixed-method assessment, while differing in whether the emphasis lies on correctness, cognitive cost, trust dynamics or reliance behavior.

    \subsection{Research Gaps and Future Directions}
    Across this taxonomy, current interfaces remain largely static and short-term. Most studies evaluate single-session interactions and fixed designs \cite{chen2025engaging, he2025conversational, kang2025themeviz, morrison2023evaluating}. There is limited understanding of how trust, reliance and expertise evolve over time. Adaptive interfaces that adjust explanation depth, interaction style and cognitive support based on user behavior and AI performance remain underexplored. Future work should also examine cross-domain generalization of interface patterns, especially in settings with ethical risk and heterogeneous stakeholders \cite{haque2024we}. Finally, achieving true human–AI collaboration requires moving beyond persuasive explanations toward designs that support critical reflection, shared control and accountable decision-making.
    

\section{Trust, Explainability and Human Confidence Calibration}
Trust and explainability play a central role in determining whether humans rely on AI systems appropriately in high-stakes decision-making. Prior work shows that neither high model accuracy nor transparent explanations alone guarantee effective human–AI collaboration. Instead, inappropriate reliance often emerges from misaligned human confidence, misunderstood AI behavior and poorly calibrated trust \cite{takayanagi2025impact, cabitza2023ai, ma2024you, inkpen2023advancing}. This taxonomy synthesizes recent work that examines how trust is formed, how explanations shape reliance and how human confidence can be calibrated to improve joint performance. The literature reveals progress in measurement and intervention design, but also highlights persistent gaps in long-term validity, personalization and interaction-aware calibration.
    \subsection{Human Confidence and Reliance Calibration}
    Several studies identify human self-confidence as a key driver of over-reliance and under-reliance on AI. High-confidence users tend to reject correct AI advice, while low-confidence users are more likely to accept incorrect recommendations \cite{takayanagi2025impact, ma2024you}. Controlled experiments show that aligning human confidence with task correctness can significantly improve human–AI team performance, even when AI accuracy remains unchanged \cite{takayanagi2025impact, ma2024you}. Feedback-based calibration mechanisms reduce under-reliance and improve final accuracy beyond human-only and AI-only baselines, though they introduce additional cognitive cost \cite{ma2024you}.

    Self-confidence shaping emerges as a promising intervention, but current approaches rely on idealized assumptions and simplified settings \cite{takayanagi2025impact}. Prediction of human confidence remains imperfect, with moderate accuracy and weak correlations between surface cues such as sentiment and actual confidence states \cite{takayanagi2025impact}. These findings suggest that confidence calibration is feasible but fragile, and that it must be supported by richer behavioral signals and adaptive mechanisms rather than static adjustments.
        
    \subsection{Explainability, Trust and Reliance Behavior}
    Explainable AI is often proposed as a solution to trust miscalibration, yet empirical results remain mixed. Explanations generally increase perceived understanding and reported trust, but they do not consistently reduce inappropriate reliance \cite{cabitza2023ai, de2025amplifying, kim2023help}. In group decision-making, explanations helped teams avoid over-reliance on incorrect AI advice and improved final accuracy, but they increased decision time and did not reduce under-reliance on correct advice \cite{de2025amplifying}. Individual users showed weaker benefits, indicating that social dynamics can amplify the effects of explainability.

    In medical and diagnostic settings, explainable AI sometimes increased over-reliance or rejection of correct AI outputs, a phenomenon described as the white-box paradox \cite{cabitza2023ai}. User-centered studies further show that explanation preferences vary widely by expertise and goals, and that users primarily value explanations that help assess reliability and limitations rather than model internals \cite{kim2023help}. Across domains, explanations function more as trust-calibration tools than as direct performance enhancers and poorly aligned explanations can worsen reliance behavior.

    \subsection{Selective Trust and Context Sensitivity}
    Trust in AI is not binary and does not depend on performance alone. Studies in personal health and logistics show that users prefer selective trust, where AI supports but does not replace human judgment \cite{van2025selective, kahr2025good}. In health decision-making, users valued factual and transparent advice but remained cautious about full reliance, especially when context or personalization was limited \cite{van2025selective}. Trust varied with individual decision-making styles, highlighting the need for adaptive and personalized interaction.

    Long-term field evidence shows that trust develops through stable mental models, workload reduction and integration into daily practice rather than short-term accuracy gains \cite{kahr2025good}. Trust was weakened by inconsistent behavior, frequent updates and limited transparency, even when performance remained acceptable. These findings contrast with short-term laboratory studies and suggest that sustainable trust depends on consistency, user control and alignment with human values over time.

    \subsection{Human Expertise, Mental Models and Complementarity}
    Effective trust calibration also depends on how AI errors interact with human expertise. Studies show that AI support benefits low-performing users regardless of tuning, while mid-performing users gain most when AI errors complement their skills \cite{inkpen2023advancing}. High-performing users rely less on AI and are better at rejecting incorrect advice, though overconfidence can reduce potential gains \cite{inkpen2023advancing}. Many users struggle to form accurate mental models of AI error patterns, which limits appropriate reliance even when explanations or confidence signals are available \cite{ma2024you, inkpen2023advancing}.

    Framework-level work formalizes these observations by distinguishing positive dominance from negative dominance and automation bias \cite{cabitza2023ai}. Causal and metric-based analyses reveal that beneficial reliance is common, but algorithmic aversion and over-reliance persist under certain explanation and task conditions. These results emphasize that trust calibration is a system-level property that emerges from the interaction between human confidence, AI behavior and interface design.

    \subsection{Datasets for Studying Human–AI Decision-Making and Confidence}
    The reviewed studies employ diverse datasets to examine confidence, trust and reliance in human–AI decision-making, but they differ in their balance between experimental control and real-world complexity. Structured datasets such as earnings call excerpts \cite{takayanagi2025impact}, mushroom classification records \cite{de2025amplifying} and Adult Income profiles \cite{ma2024you} rely on careful preprocessing, balanced labels and calibrated or intentionally perturbed model accuracy to isolate self-confidence shaping and over-reliance effects. These designs allow precise causal analysis but simplify task context. In contrast, medical and citizen science datasets introduce higher decision stakes and variability through expert-grounded annotations and difficulty characterization, as shown in multi-modal clinical cases with staged AI and XAI support \cite{cabitza2023ai} and complex video-based vessel classification in Stall Catchers \cite{inkpen2023advancing}. Mixed-method and qualitative approaches further extend this landscape by prioritizing ecological validity over dataset uniformity. Health decision-making with MoveAI \cite{van2025selective} and long-term logistics planning interviews \cite{kahr2025good} rely on participant interactions and thematic analysis to capture trust dynamics that evolve over time. Real-world AI use is also approximated through realistic yet controlled settings, such as mock explainable outputs in bird identification \cite{kim2023help}, which decouple explanation design from proprietary models. Across these works, dataset choices reflect a trade-off between controllability and realism, suggesting that observed patterns of confidence, trust and reliance are tightly coupled to how data are selected, structured and presented to human decision-makers.

    \subsection{Evaluation Metrics for Human–AI Decision-Making}
    The reviewed studies adopt heterogeneous evaluation matrices that reflect different assumptions about what constitutes effective and trustworthy human–AI decision-making. Controlled experiments rely on behavioral and performance-based metrics to quantify reliance and confidence alignment, as seen in acceptance rate and under- and over-reliance measures \cite{takayanagi2025impact}, reliance typologies and decision accuracy in group settings \cite{de2025amplifying}, and detailed confidence calibration and reliance appropriateness metrics such as ECE and confidence–correctness matching \cite{ma2024you}. Clinical and high-stakes domains extend these approaches by introducing dominance- and safety-oriented measures, including Weight of Advice, automation bias and causal estimates of AI effectiveness that distinguish correct and incorrect AI influence \cite{cabitza2023ai}. In contrast, health and logistics studies emphasize perception- and trust-oriented evaluations through statistical summaries, reliability testing and thematic analysis, rather than direct performance metrics \cite{van2025selective, kahr2025good}. Real-world and exploratory contexts further reduce reliance on numerical matrices, favoring qualitative coding and interpretive analysis to capture user needs and explanation use, as demonstrated in bird identification with mock XAI outputs \cite{kim2023help}. Hybrid designs integrate both objective and subjective assessments, combining accuracy, tuning effects and user perception scales to study team performance and trust calibration under varying AI behaviors \cite{inkpen2023advancing}. Across these works, evaluation choices reveal a trade-off between metric precision and ecological validity, indicating that conclusions about confidence, reliance and trust are strongly shaped by the selected evaluation framework rather than by model performance alone.

    \subsection{Research Gaps and Future Directions}
    Current research remains limited by short-term studies, static assumptions about users and simplified decision settings \cite{takayanagi2025impact, cabitza2023ai, de2025amplifying, ma2024you}. There is limited understanding of how confidence, trust and reliance evolve over time or across changing task conditions. Future work should focus on adaptive calibration strategies that jointly model human confidence and AI uncertainty, while accounting for expertise, cognitive cost and user experience. Longitudinal and real-world evaluations are needed to validate calibration mechanisms beyond controlled experiments \cite{kahr2025good}. Finally, explainability should move beyond generic transparency toward context-sensitive, user-informed designs that support accurate mental models and selective trust rather than persuasion or compliance.

\section{Collaborative Decision Frameworks \& Optimization of Human–AI Teams}
Human–AI collaboration has moved beyond static decision aids toward interactive frameworks that aim to optimize joint performance under uncertainty, time pressure and cognitive constraints. Prior work shows that fully automated systems perform poorly in novel or ambiguous settings, while naïve AI assistance can increase overreliance and miscalibrated trust \cite{tariq2025a2c, buccinca2024towards, swaroop2024accuracy}. This taxonomy reviews recent approaches that study and optimize collaborative decision-making by examining interaction dynamics, adaptive assistance and system-level coordination. The literature reflects progress in modeling collaboration and human factors, but also exposes gaps in scalability, realism and long-term optimization of human–AI teams.
    \subsection{Interaction-Aware Analysis of Collaborative Decision Making}
    Understanding how humans combine multiple AI models is a prerequisite for effective collaboration. Retrospective End-User Walkthroughs provide a structured method to analyze decision rationales, trust formation and cognitive load in complex AI-supported systems \cite{de2023retrospective}. Empirical results show that AI support improves decision accuracy but increases task completion time, highlighting a clear accuracy–effort trade-off. Many users relied primarily on visual analytics and underused AI models due to poor visibility, system latency or unclear integration \cite{de2023retrospective}. Trust depended on transparency and explainability, while confirmation bias shaped how recommendations were accepted or ignored.

    These findings align with controlled studies showing that human reliance patterns are often stable and predictable early in interaction \cite{swaroop2024accuracy}. Both strands of work reveal that collaboration failures often stem from interface design and workflow placement rather than model accuracy alone \cite{de2023retrospective, swaroop2024accuracy}. However, reliance on recall and simplified lab tasks limits understanding of long-term adaptation and real-world complexity. Future work should combine interaction-aware analysis with online behavioral signals to reduce recall bias and better support dynamic collaboration.
    
    \subsection{Framework-Level Coordination Under Uncertainty}
    Several studies propose structured frameworks to coordinate automation, deferral and collaboration. The A²C framework formalizes collaboration through a modular pipeline that separates confident automation, human deferral and collaborative exploration \cite{tariq2025a2c}. Results show that collaborative exploration consistently outperforms full automation and simple deferral, especially in novel or unknown-class settings. These gains arise from combining human reasoning with AI uncertainty awareness, rather than replacing human judgment.

    Time-pressure studies provide complementary evidence that fixed assistance strategies are suboptimal \cite{swaroop2024accuracy}. Under time pressure, AI-before assistance accelerates decisions but amplifies overreliance, while AI-after assistance supports correction on difficult tasks. Both works indicate that collaboration must adapt to uncertainty, task difficulty and human behavior to achieve complementarity \cite{tariq2025a2c, swaroop2024accuracy}. Despite their strengths, these frameworks rely heavily on simulations or simplified tasks, and they offer limited insight into how collaboration scales to complex, high-stakes environments with latency, privacy or reliability constraints.

    \subsection{Adaptive Optimization of Human–AI Team Performance}
    Beyond static frameworks, adaptive optimization approaches treat human–AI interaction as a sequential decision problem. Offline reinforcement learning has been used to learn assistance policies that optimize not only accuracy but also human-centric objectives such as learning and engagement \cite{buccinca2024towards}. Accuracy-optimized policies improve performance and reduce harmful overreliance, while learning-optimized policies tailor assistance based on individual traits such as Need for Cognition. These results suggest that adaptive assistance can regulate reliance without suppressing engagement.

    Time-pressure experiments reinforce the need for such personalization. Overreliance emerges consistently across individuals and can be predicted early, enabling adaptive strategies that match assistance timing to user behavior and task difficulty \cite{swaroop2024accuracy}. However, learning gains remain difficult to achieve and depend on strong learning signals and effective explanation design \cite{buccinca2024towards}. Current optimization methods also assume simplified uncertainty models and limited task diversity, which constrains generalization.

    \subsection{Datasets for Human–AI Decision-Making Systems}
    Several studies construct datasets to study how humans interact with multiple AI models under uncertainty, but they differ in task domains, data composition and evaluation goals. One line of work focuses on complex real-world decision settings. The financial decision-making dataset in \cite{de2023retrospective} combines four years of structured financial data for 268 companies with unstructured external news, enabling analysis of how users interpret predictions from multiple AI models with different training attributes. In contrast, \cite{buccinca2024towards} and \cite{swaroop2024accuracy} rely on controlled human-subject datasets to isolate behavioral effects. The dataset in \cite{buccinca2024towards} captures sequential human decisions in an exercise prescription task and records interactions with different AI assistance policies, including deliberately incorrect recommendations, to study learning and accuracy trade-offs. Similarly, \cite{swaroop2024accuracy} generates a synthetic but carefully controlled logic-puzzle dataset that simulates medical prescriptions under time pressure, allowing precise measurement of overreliance and response behavior. While these datasets emphasize human behavior, \cite{tariq2025a2c} focuses on uncertainty handling in AI-assisted classification. It constructs benchmark datasets by systematically splitting known and unknown classes across vision and security domains, enabling evaluation of human–AI collaboration under novelty. Compared to \cite{de2023retrospective}, which emphasizes model diversity and interpretability in realistic analytics systems, \cite{tariq2025a2c} prioritizes formal dataset partitioning to test robustness to unseen classes. Together, these datasets highlight a trade-off between ecological validity and experimental control, and they provide complementary foundations for evaluating human–AI decision-making systems across domains.

    \subsection{Evaluation Metrics for Human–AI Decision-Making}
    Across these studies, evaluation metrics reflect different priorities in assessing human–AI decision-making, balancing task performance, collaboration quality and human-centric outcomes. Task-level accuracy is a common baseline, but it is operationalized differently. In \cite{de2023retrospective}, correctness and task completion time are used to contrast AI-supported and non-AI workflows, revealing a trade-off between higher accuracy and increased time cost. In contrast, \cite{swaroop2024accuracy} refines accuracy into graded scores to capture suboptimal but acceptable decisions and pairs it with response time to study behavior under time pressure. Beyond raw performance, several works introduce metrics that explicitly measure collaboration. The A²C framework in \cite{tariq2025a2c} uses F1-score to compare automation and deferral modes, and proposes the CoEx Success Rate to quantify how effectively humans and AI resolve uncertain cases together. Human-centric effects are further emphasized in \cite{buccinca2024towards}, which evaluates learning, overreliance and subjective perceptions such as trust and mental demand alongside objective accuracy. Qualitative measures in \cite{de2023retrospective}, including think-aloud data and user experience scores, complement quantitative metrics by revealing interaction strategies and cognitive effort. Taken together, these metrics show a shift from single-score accuracy toward multi-dimensional evaluation that captures efficiency, collaboration and human reliance in AI-assisted decision-making systems.

    \subsection{Research Gaps and Future Directions}
    Existing work demonstrates that collaborative decision frameworks outperform automation alone, but several gaps remain. First, most studies rely on short-term tasks, simulations or controlled experiments, limiting external validity \cite{de2023retrospective, tariq2025a2c, buccinca2024towards, swaroop2024accuracy}. Second, optimization focuses primarily on accuracy and reliance, while long-term learning, trust evolution and cognitive cost remain underexplored \cite{de2023retrospective, buccinca2024towards}. Third, integration challenges such as information architecture, AI visibility and workflow alignment continue to hinder effective collaboration \cite{de2023retrospective, swaroop2024accuracy}.

    Future research should develop adaptive frameworks that jointly model uncertainty, time pressure and individual differences in real-world settings. Longitudinal evaluations are needed to assess how human–AI teams co-adapt over time. Finally, optimization objectives should expand beyond performance to include sustainable trust, interpretability and ethical personalization, enabling robust and effective human–AI collaboration across domains.

\section{Human-AI Interaction in High-Stakes Healthcare Decision Support Systems}

Human-AI interaction research has increasingly examined empirical interaction patterns that emerge when AI systems are deployed in high-stakes healthcare decision support systems. Such settings have imposed strict constraints related to patient safety, professional accountability, and ethical responsibility. Medical decision-making has involved uncertainty, incomplete information, and irreversible consequences. Empirical investigations have analyzed clinician behavior, reliance strategies, and interaction dynamics when AI systems have supported diagnosis, treatment selection, and clinical model refinement \citep{wang2024llmcdss, zhang2023negotiate}. Evidence from these evaluations has indicated that interaction design, trust calibration, and contextual alignment have played a central role in shaping effective human-AI collaboration \citep{burgess2023design}.

Healthcare decision support systems have differed from low-risk applications due to regulatory oversight and clinician liability. As a result, AI systems have been assessed not only for predictive performance but also for interpretability, reliability, and compatibility with established clinical workflows \citep{li2024oncology}. These characteristics have positioned healthcare as a critical domain for studying human-AI interaction under high-stakes conditions.

\subsection{LLM-Augmented Clinical Decision Support}

Recent empirical work has examined the integration of large language models into clinical decision support systems \citep{wang2024llmcdss}. These systems have been designed to assist clinicians with guideline interpretation and contextual reasoning during time-sensitive clinical tasks. Conversational interfaces have been embedded within broader decision support frameworks that have also included predictive models and visual dashboards.

Empirical evaluations have shown that clinicians have primarily used LLM-based components as reference tools rather than as decision authorities. Clinicians have queried these systems to clarify clinical guidelines, patient risk factors, and recommended actions. Reliance has depended strongly on transparency and evidence attribution. When system responses have lacked clear provenance or clinical justification, clinicians have expressed skepticism and have reduced reliance.

Interaction differences have emerged across expertise levels. Less experienced clinicians have perceived LLM-based systems as sources of expert guidance. More experienced clinicians have used the same systems to validate or challenge their own judgment. These findings have indicated that LLM-augmented decision support must accommodate heterogeneous user expertise within healthcare teams.

\subsection{Clinician Interaction with AI-Based Treatment Recommendations}

Empirical studies have examined clinician interaction with AI-based treatment recommendation systems in critical care environments \citep{zhang2023negotiate}. These systems have supported treatment decisions under time pressure and clinical uncertainty. Recommendations have been presented through interpretable and interactive visualization interfaces designed to fit existing clinical workflows.

Observed clinician behavior has departed from binary acceptance models. Clinicians have not simply followed or rejected AI recommendations. Instead, multiple interaction strategies have been observed. Some clinicians have ignored recommendations when bedside assessments conflicted with model outputs. Others have trusted recommendations when they aligned with clinical expectations. Many clinicians have selectively adopted parts of recommendations while modifying others. This behavior has reflected attempts to integrate algorithmic guidance with contextual patient information.

Explainability mechanisms have influenced clinician confidence but have not guaranteed acceptance. Visual explanations have supported understanding of model reasoning. However, explanations have increased cognitive effort in some cases. These observations have suggested that explanation design must be sensitive to clinical context and task urgency.

\subsection{Dynamic Human-AI Interaction in Precision Oncology}

Human-AI interaction has become more complex in medical domains that require repeated decisions across extended treatment timelines. Research in precision oncology has examined AI-assisted decision-making across evolving patient trajectories \citep{li2024oncology}. In these settings, clinicians have adjusted treatment strategies based on disease progression, patient response, and anticipated outcomes.

Empirical findings have shown that trust in AI has evolved over time rather than remaining static. Initial skepticism has decreased when AI recommendations aligned with clinical goals. Trust has declined when outputs conflicted with patient-specific context or clinical intuition. Interfaces that have included uncertainty visualization and trade-off representations have supported reflective decision-making.

\begin{table*}[t]
\caption{Comparison of empirical studies examining whether AI functions as a teammate or a tool through human–AI interaction patterns, interface designs and empirical outcomes across decision-critical domains, including high-stakes healthcare decision support.}
\centering
\small
\setlength{\tabcolsep}{4pt}
\begin{tabular}{|p{0.14\textwidth} p{0.20\textwidth} p{0.16\textwidth} p{0.22\textwidth} p{0.22\textwidth}|}
\hline
\textbf{Study} & \textbf{Decision Context} & \textbf{AI Role} & \textbf{Interaction Pattern} & \textbf{Core Empirical Insight} \\
\hline
{\cite{de2023retrospective}} & Financial analytics & Multi-model decision aid & Visual analytics with walkthrough & Accuracy increased but effort and time also increased. \\
\hline
\citep{burgess2023design} & Treatment planning & Insight generation & Clinician reflection & Adoption has depended on epistemic alignment. \\
\hline
{\cite{cabitza2023ai}} & Medical diagnosis & Decision influence analysis & Causal dominance assessment & Explainability altered dominance and reliance patterns. \\
\hline
{\cite{kim2023help}} & Wildlife identification & Reliability assessment & User-informed explanations & Explanations supported trust calibration and learning. \\
\hline
{\cite{inkpen2023advancing}} & Citizen science & Decision assistance & Expertise-aware AI tuning & Complementarity depended on user skill alignment. \\
\hline
{\cite{morrison2023evaluating}} & Disaster assessment & Visual classification & Human-like explanations & Causal explanations reduced over-reliance when accurate. \\
\hline
\citep{zhang2023negotiate} & ICU treatment & Recommendation support & Selective trust and negotiation & Clinicians have negotiated with AI advice. \\
\hline
\citep{wang2024llmcdss} & Clinical decision support & Guideline reasoning aid & Conversational querying & Trust has depended on evidence grounding. \\
\hline
{\cite{buccinca2024towards}} & Behavioral decision-making & Adaptive assistance & RL-driven policy selection & Personalization reduced over-reliance and improved outcomes. \\
\hline
{\cite{haque2024we}} & Predictive policing & Risk analysis aid & Stakeholder-centered interaction & Trust depended on lived experience and ethics. \\
\hline
{\cite{ma2024you}} & Income prediction & Confidence calibration & Feedback-based adjustment & Confidence alignment improved final accuracy. \\
\hline
{\cite{swaroop2024accuracy}} & Time-critical decisions & Decision acceleration & Timing-based AI support & Time pressure amplified reliance on AI advice. \\
\hline
\citep{li2024oncology} & Precision oncology & Sequential decision support & Dynamic interaction & Trust has evolved across repeated use. \\
\hline
{\cite{kahr2025good}} & Logistics planning & Operational support & Long-term human-AI use & Trust emerged from stability and workflow integration. \\
\hline
{\cite{kang2025themeviz}} & Qualitative analysis & Theme suggestion & Visual sensemaking with LLMs & Structure reduced load but limited collaboration. \\
\hline
{\cite{takayanagi2025impact}} & Financial investment & Confidence shaping & Confidence-aware intervention & Aligning confidence improved appropriate reliance. \\
\hline
{\cite{de2025amplifying}} & Group classification & Decision explanation & Team-based XAI use & Explanations reduced over-reliance in groups. \\
\hline
{\cite{van2025selective}} & Personal health & Advisory support & Shared decision-making dialogue & Users preferred selective trust over delegation. \\
\hline
{\cite{tariq2025a2c}} & High-risk classification & Automation coordinator & Deferral and collaboration & Collaboration outperformed automation under uncertainty. \\
\hline
\citep{gao2025adversarial} & High-stakes analysis & Analytical support & Human-in-the-loop reliance & Robustness has depended on human behavior. \\
\hline
{\cite{chen2025engaging}} & Health planning & Recommendation support & Explanation-focused interfaces & Transparency improved trust but increased over-reliance. \\
\hline
{\cite{cho2025human}} & Surgical verification & Quality assurance & Human+XAI review & Explainability improved precision without added workload. \\
\hline
\citep{chen2025medebiaser} & Medical imaging & Bias mitigation & Human feedback loops & Direct clinician input has reduced bias. \\
\hline
{\cite{he2025conversational}} & Financial approval & Decision explanation & Conversational XAI & Fluency increased confidence without better calibration. \\
\hline
\end{tabular}

\label{tab:human_ai_interaction_all}
\end{table*}
\clearpage

Both over-reliance and under-reliance have resulted in suboptimal decisions. Excessive skepticism has led clinicians to dismiss useful recommendations. Excessive deference has reduced critical evaluation in complex cases. These findings have emphasized the need for decision support systems that promote calibrated reliance.

\subsection{Human-in-the-Loop Bias Mitigation in Medical AI}

Bias has remained a significant concern in healthcare AI applications. Interactive human-in-the-loop systems have enabled clinicians to participate directly in bias identification and mitigation during model development \citep{chen2025medebiaser}. These systems have allowed clinicians to inspect predictions, analyze explanation outputs, and provide corrective feedback through structured interfaces.

User studies have shown that clinicians have identified biases related to label imbalance and co-occurrence patterns in medical data. Visual inspection tools have supported targeted correction and error analysis. This interaction has reduced communication delays between clinicians and engineers and has improved collaboration efficiency.

\subsection{Design Considerations for Healthcare Decision Support}

Empirical investigations based on clinician interviews have derived design considerations for healthcare treatment decision support systems \citep{burgess2023design}. These studies have shown that clinicians have evaluated AI tools based on methodological rigor, alignment with clinical evidence, and integration with established workflows.

Trust has been grounded in how recommendations have been generated rather than how outputs have been presented. Clinicians have compared AI-derived insights with established evidence sources such as clinical guidelines and randomized trials. Support for clinician autonomy has remained central. Systems have been more acceptable when they have encouraged reflection and deliberation rather than enforced compliance.

\subsection{Human Factors Influencing Reliance and Decision Authority}

Across the examined empirical studies, human factors have played a decisive role in shaping how decision authority has been distributed between clinicians and AI systems in high-stakes healthcare decision support. Clinician expertise, perceived clinical risk, and task urgency have consistently influenced reliance strategies. More experienced clinicians have demonstrated selective and conditional use of AI outputs, while less experienced clinicians have relied more heavily on system guidance.

Empirical evidence has indicated that decision authority has rarely been fully delegated to AI systems. Clinicians have retained final responsibility for decisions and have overridden AI recommendations when outputs conflicted with bedside assessment or patient-specific context. These patterns have suggested that human-AI interaction in healthcare has been characterized by negotiated authority rather than automation-driven delegation.

\subsection{Implications for Human-AI Collaboration in High-Stakes Decision Support}

Evidence from empirical evaluations has indicated several implications for human-AI collaboration in high-stakes healthcare decision support. AI systems have functioned primarily as supportive collaborators that inform, validate, or challenge clinician judgment rather than as autonomous decision-makers.

Effective collaboration has depended on interaction mechanisms that support contextual reasoning instead of prescriptive decision-making. Systems that have enabled uncertainty assessment, evidence inspection, and exploration of alternatives have supported more effective collaboration. These findings have suggested that adaptive interaction designs are necessary to accommodate variation in clinician expertise, decision urgency, and clinical risk.

\section{Key Findings and Trending Research Directions}
Across the reviewed literature, a consistent pattern emerges that effective human–AI interaction depends more on interface alignment than on model capability alone. Explanation richness, conversational fluency and visual detail often increase perceived trust and understanding, but they do not reliably improve decision quality. In several domains, especially high-stakes and time-sensitive tasks, fluent explanations and conversational interfaces amplified over-reliance and reduced critical judgment. In contrast, interfaces grounded in task-relevant signals, uncertainty cues and workflow-compatible visualizations supported better precision and corrective behavior without increasing cognitive burden. These findings suggest that explanation effectiveness is conditional and context-dependent, and that poorly aligned transparency can be harmful rather than beneficial.

Trust calibration remains a central challenge that cuts across interaction design, explainability and collaboration frameworks. The evidence shows that trust is shaped not only by interface transparency but also by user expertise, lived experience and ethical concerns. Mechanisms that expose AI uncertainty or confidence can support calibration, yet they often increase cognitive effort and slow decision-making. Studies in healthcare, policing and finance further demonstrate that trust develops through consistent behavior, integration into existing practices and respect for human judgment, rather than through persuasive explanations. Short-term laboratory gains in trust or confidence often fail to translate into sustained or appropriate reliance in real-world settings.

Finally, prior work reveals a gap between current interaction designs and the needs of long-term adaptive human–AI collaboration, as static interfaces and single-session studies limit understanding of evolving trust, reliance and expertise, while adaptive frameworks remain confined to simplified, controlled settings. Future research should prioritize longitudinal studies, adaptive interface strategies and ethically grounded designs that support selective trust, shared control and accountable decision-making. Without such advances, human–AI systems risk optimizing short-term performance at the expense of robust and sustainable collaboration. 
\newline
Table~\ref{tab:human_ai_interaction_all} contrasts human–AI interaction patterns across decision-critical domains.

\section{Conclusion}
The analysis indicates that the primary barrier to AI becoming a teammate is not computational power, but rather interaction rigidity. Highly fluent interfaces, such as conversational agents, often simulate the experience of a teammate without delivering the same level of reliability, leading to dangerous over-reliance. In contrast, successful implementations in healthcare and logistics demonstrate that treating AI as a context-aware tool that respects human authority and supports selective trust yields more effective outcomes.
Achieving a true partnership necessitates a paradigm shift from optimizing model accuracy to optimizing the Human-AI ecosystem. Future research should focus on longitudinal studies tracking the evolution of trust, adaptive interfaces that adjust behaviour based on user confidence, and ethical frameworks that ensure accountability. Until systems can actively model human partners and adapt to dynamic contexts, AI will remain an underutilised yet powerful tool, rather than a true teammate.

\section*{Limitations}
This review has several limitations. First, our analysis is based on a synthesis of existing HAI literature and therefore depends on the conceptual frameworks, evaluation methods and assumptions adopted in prior studies. As a result, empirical evidence on long-term human–AI collaboration, especially from real-world deployments, remains limited. Second, while we include healthcare applications to ground the discussion, our coverage of other high-stakes domains is selective and may not fully capture domain-specific constraints or practices. Third, much of the reviewed work relies on short-term user studies or controlled experimental settings, which makes it difficult to assess how trust calibration, authority negotiation and shared mental models evolve over time. Finally, our proposed dimensions and interpretations of “tool” versus “teammate” involve subjective judgments, and alternative taxonomies or theoretical lenses may lead to different conclusions.

\section*{Acknowledgments}
This study was conducted collaboratively to review human-AI interaction in decision support systems, with each author contributing equally.


\bibliography{custom}

\end{document}